\documentstyle[12pt,amsmath,amssymb,subeqnarray,cite,times]{article}
 \def\ml{m_\ell}
 \def\N{N}                       
 \def\Q{{\mathcal Q}}
 \def\lpm{\eta\strut_{\pm\lambda}\eta\strut_{\pm\lambda'}}
 \def\lmp{\eta\strut_{\mp\lambda}\eta\strut_{\mp\lambda'}}
 \def\llp{\eta\strut_{ \lambda\lambda'}}
 \def\llm{\eta\strut_{-\lambda\lambda'}}
\begin{document}

\begin{center}
{\bfseries LEPTON POLARIZATION IN NEUTRINO--NUCLEON INTERACTIONS}
\vskip 5mm
{\bf Konstantin\,S.\,Kuzmin},$^{1,2}$
{\bf Vladimir\,V.\,Lyubushkin}$^{1,3}$
and 
{\bf Vadim\,A.\,Naumov}$^{1,4}$
\vskip 3mm
{\small
$^{(1)}$ {\small\em Joint Institute for Nuclear Research, Dubna, Russia}
\\
$^{(2)}$ {\small\em Institute for Theoretical and Experimental Physics,
         Moscow, Russia}
\\
$^{(3)}$ {\small\em Physics Department of Irkutsk State University,
         Irkutsk, Russia}
\\
$^{(4)}$ {\small\em Dipartimento di Fisica and INFN, Sezione di Firenze,
         Sesto Fiorentino, Italy}
}
\end{center}
\vskip 5mm

\begin{abstract}
We derive generic formulas for the polarization density matrix
of leptons produced in ${\nu}N$ and $\overline{\nu}N$ collisions
and briefly consider some important particular cases.
Next we employ the general formalism in order to include the
final lepton mass and spin into the popular model by Rein and
Sehgal for single pion neutrinoproduction.
\end{abstract}

\vskip 8mm

\section{Introduction}
\label{sec:Introduction}

Polarization of leptons generated in ${\nu}N$ and $\overline{\nu}N$
collisions is important for studying neutrino oscillations and relevant
phenomena in experiments with atmospheric and accelerator neutrino beams.
Let us shortly touch upon a few illustrative examples.

$\bullet$ Contained $\tau$ lepton events provide the primary
signature for $\nu_\mu\to\nu_\tau$ oscillations. Besides they are a
source of unavoidable background to the future proton decay experiments.
But a low or intermediate energy $\tau$ lepton generated inside a water
Cherenkov detector is unobservable in itself and may only be identified
through the $\tau$ decay secondaries whose momentum configuration is
determined by the $\tau$ lepton helicity.

$\bullet$ In case of $\nu_\mu-\nu_\tau$ mixing, the leptonic decays
of $\tau$'s generated inside the Earth yield an extra contribution
into the flux of through-going upward-going muons (TUM) and stopping
muons (SM) which is absent in case of $\nu_\mu-\nu_s$ or $\nu_\mu-\nu_e$
mixing. The absolute value and energy spectrum of the
``$\tau_{\mu3}$ muons'' are affected by the $\tau$ beam polarization.
The contribution is evidently small but measurable in future
large-scale experiments, particularly those with magnetized tracking
calorimeters (like in the experiments NuMI--MINOS and MONOLITH).
Note that the energy and angular distributions of the charge ratio for
the ``$\tau_{\mu3}$ muons'' are considerably differ from those for the
``direct'' TUM and SM since the longitudinal polarizations of $\tau^+$
and $\tau^-$ have in average opposite signs.

$\bullet$ Decay of $\nu_\mu$ or $\overline{\nu}_\mu$ induced muons
with energy below the detection threshold may produce detectable
electrons whose energy distributions are affected by the muon
polarization. Such events, being classified as ``$e$-like''
(for a water detector) or ``showering'' (for an iron detector),
mimic the $\nu_e$ or $\overline{\nu}_e$ induced events.

In this paper, we derive general formulas for the lepton
polarization density matrix by applying a covariant method
(Sect.~\ref{sec:PDM}) and briefly consider their applications to
deep inelastic, quasielastic and resonance neutrino interactions.
We explicitly demonstrate that the perpendicular and transverse
polarizations are dependent of an intrinsically indeterminate
phase and thus unobservable in contrast with the longitudinal
polarization and degree of polarization.
In Sect.~\ref{sec:RS} we discuss with some details a generalization
of the Rein--Sehgal model for single pion neutrinoproduction through
baryon resonances which takes into account the final lepton mass
and spin.

\section{Polarization density matrix}
\label{sec:PDM}

The lepton polarization vector $\boldsymbol{\mathcal{P}}=
\left({\mathcal P}_P,{\mathcal P}_T,{\mathcal P}_L\right)$
is defined through the polarization density matrix
$\boldsymbol{\rho}=\frac{1}{2}\left(1+\boldsymbol{\sigma}
\boldsymbol{\mathcal{P}}\right)$ whose matrix elements are given
by contracting the leptonic tensor $L^{\alpha\beta}_{\lambda\lambda'}$
with the spin-averaged hadronic tensor $W_{\alpha\beta}$.
The leptonic tensor is given by
\begin{equation}\label{LeptonicTensor}
L_{\lambda\lambda'}^{\alpha\beta}=\left\{
\begin{aligned}
j^{\,\alpha}_{\lambda}\left(j^{\,\beta}_{\lambda'}\right)^*
\enskip\text{with}\enskip
j\strut^{\,\alpha}_\lambda
=\overline{u}\left(k',s\right)\gamma^\alpha
\left(\frac{1-\gamma_5}{2}\right)u(k)
&\enskip\text{for}\enskip\nu_\ell, \\
\overline{j}^{\,\alpha}_{\lambda}
\left(\overline{j}^{\,\beta}_{\lambda'}\right)^*
\enskip\text{with}\enskip
\overline{j}\strut^{\,\alpha}_\lambda
= \overline{v}(k)\gamma^\alpha
\left(\frac{1-\gamma_5}{2}\right)v\left(k',s\right)
&\enskip\text{for}\enskip
\overline{\nu}_\ell,
\end{aligned}
\right.
\end{equation}
where $k$ and $k'$ are the 4-momenta of $\nu_\ell$
or $\overline{\nu}_\ell$ and lepton $\ell^-$ or $\ell^+$
($\ell=e,\mu,\tau$), $\lambda$ and $\lambda'$ are the
lepton helicities, $s$ is the axial 4-vector of the
lepton spin.

It can be shown that the weak leptonic currents
$j\strut^{\,\alpha}_\lambda$ and $\overline{j}\strut^{\,\alpha}_\lambda$
are expressed as
\begin{subequations}\label{CC}
\begin{align}
\label{CC_a}
j\strut^{\,\alpha}_\lambda
& = \phantom{\lambda}N_\lambda\left[k'^{\alpha}(ks)-s^{\alpha}(kk')
-i\epsilon^{\alpha\beta\gamma\delta}
  s_{\beta}k_{\gamma}k'_{\delta}+\ml k^{\alpha}\right],         \\
\label{CC_b}
\overline{j}\strut^{\,\alpha}_\lambda
& ={\lambda}N_\lambda\left[k'^{\alpha}(ks)
 -s^{\alpha}(kk')+i\epsilon^{\alpha\beta\gamma\delta}
  s_{\beta}k_{\gamma}k'_{\delta}-\ml k^{\alpha}\right].
\end{align}
\end{subequations}
Here $\ml$ is the lepton mass and the normalization constant
$N_\lambda$ is expressed in terms of the kinematic variables
and of two intrinsically indeterminate phases $\varphi_+$ and
$\varphi_-$:
\begin{equation}\label{N_lambda}
N_\lambda = \frac{\exp\left(\pm i\varphi_\lambda\right)}
            {\sqrt{\left(kk'\right)\pm\ml(ks)}}
          = \frac{\exp\left(\pm i\varphi_\lambda\right)}{\ml}
            \sqrt{\frac{E_\ell \mp\lambda P_\ell}
            {E_\nu\left(1\mp\lambda\cos\theta\right)}}.
\end{equation}
Here $E_\nu$, $E_\ell$ and $\theta$ are, respectively, the
incident neutrino energy, lepton energy and the scattering
angle of the lepton in the lab.\ frame,
$P_\ell=\sqrt{E_\ell-\ml^2}$;
the upper (lower) signs are for $\nu_\ell$ ($\overline{\nu}_\ell$).
As it follows from Eq.~\eqref{CC_a}, the components of the neutrino
current are
\begin{equation}\label{j_components}
\begin{aligned}
j^0_\lambda
& =\exp\left(i\varphi_\lambda\right)
\sqrt{\left(1-\lambda\cos\theta\right)E_\nu
\left(E_\ell-\lambda P_\ell\right)}, \\
\mathbf{j}_\lambda
& =\frac{\lambda\ml N_\lambda}{P_\ell}
\left[\lambda P_\ell\mathbf{k}-E_\nu\mathbf{k}'
+i\left(\mathbf{k}\times\mathbf{k}'\right)\right].
\end{aligned}
\end{equation}
As it follows from Eqs.~\eqref{CC} and \eqref{N_lambda}, the currents
for neutrino and antineutrino are related by
\begin{equation}\label{Nu-Nubar_relation}
\overline{j}^{\,\alpha}_\lambda=
-\lambda\left(j^{\alpha}_{-\lambda}\right)^*.
\end{equation}

We use the standard representation for the hadronic tensor
(see, e.g., Ref.~\cite{LlewellynSmith:72})
\begin{align}\label{HadronicTensor}
W_{\alpha\beta}
= &\ - g_{\alpha\beta}\,W_1
     +   \frac{p_\alpha\,p_\beta}{M^2}\,W_2
     -   \frac{i\,\epsilon_{\alpha\beta\rho\sigma}\,
               p^\rho\,q^\sigma}{2M^2}\,W_3  \nonumber\\
  &\ +   \frac{q_\alpha\,q_\beta}{M^2}\,W_4
     +   \frac{p_\alpha\,q_\beta+q_\alpha\,p_\beta}{2M^2}\,W_5
     +i\,\frac{p_\alpha\,q_\beta-q_\alpha\,p_\beta}{2M^2}\,W_6
\end{align}
which includes 6 nucleon structure functions, $W_n$,
whose explicit form is defined by the particular subprocess
(QE, RES or DIS). Here $p$ and $M$ are the nucleon 4-momentum
and mass, respectively, $q=k-k'$ is the $W$ boson
4-momentum. By applying Eqs.~\eqref{LeptonicTensor}, \eqref{CC}
and \eqref{HadronicTensor}, we obtain
\begin{equation*}
\rho_{\lambda\lambda'}
=\frac{L_{\lambda\lambda'}^{\alpha\beta}W_{\alpha\beta}}
      {4ME_\nu\mathcal{R}}
=\frac{\ml^2E_{\nu}N_\lambda N_{\lambda'}^*}
      {4M\mathcal{R}}\sum_{k=1}^6A_{\lambda\lambda'}^kW_k,
\end{equation*}
\begin{align*}
A_{\lambda\lambda'}^1
=&\ 2\left(\llp\mp\llm\right)\sin^2\theta,                         \\
A_{\lambda\lambda'}^2
=&\ 4\left(\lpm\sin^4\frac{\theta}{2}
    +\lmp\cos^4\frac{\theta}{2}\right)
     \pm\llm\sin^2\theta, \\
A_{\lambda\lambda'}^3
=&\ \pm\sin^2\theta\left(\lpm\frac{E_\nu-P_\ell}{M}
   +\lmp\frac{E_\nu+P_\ell}{M}\mp\llm\frac{E_\nu}{M}\right),       \\
A_{\lambda\lambda'}^4
=&\ 4\left[\lpm\frac{(E_\nu+P_\ell)^2}{M^2}\sin^4\frac{\theta}{2}
    +\lmp\frac{(E_\nu-P_\ell)^2}{M^2}\cos^4\frac{\theta}{2}\right] \\
 &\  \pm\llm\frac{\ml^2}{M^2}\sin^2\theta,                         \\
A_{\lambda\lambda'}^5
=&\-4\left[\lpm\frac{E_\nu+P_\ell}{M}\sin^4\frac{\theta}{2}
    +\lmp\frac{E_\nu-P_\ell}{M}\cos^4\frac{\theta}{2}\right]       \\
 &\  \mp\llm\frac{E_\ell}{M}\sin^2\theta,                          \\
A_{\lambda\lambda'}^6
=&\ i\left(\frac{\lambda-\lambda'}{2}\right)\frac{P_\ell}{M}\sin^2\theta,
\end{align*}
where $\eta_\lambda\equiv(1+\lambda)/2$ ($\eta_+=1$, $\eta_-=0$)
and the dimensionless normalization factor
$\mathcal{R}$ is given by the condition Tr\,$\boldsymbol{\rho}=1$:
\begin{align*}
\mathcal{R}=&\  \frac{1}{4ME_\nu}
                \left(L_{++}^{\alpha\beta}
                     +L_{--}^{\alpha\beta}\right)W_{\alpha\beta} \\
           =&\ \left(\frac{E_\ell-P_\ell\cos\theta}{ M}\right)
                  \left(W_1+\frac{\ml^2}{2M^2}W_4\right)
                + \left(\frac{E_\ell+P_\ell\cos\theta}{2M}\right)W_2 \\
            &\ \pm\left[\left(\frac{E_\nu+E_\ell}{M}\right)
                  \left(\frac{E_\ell-P_\ell\cos\theta}{2M}\right)
                - \frac{\ml^2}{2M^2}\right]W_3-\frac{\ml^2}{2M^2}W_5.
\end{align*}
Now we can find the explicit formulas for the elements of the polarization
density matrix in terms of variables $E_\nu$, $P_\ell$ and $\theta$:
\begin{align*}
  \rho_{++}\left(E_\nu,P_\ell,\theta\right)
&=\rho_{--}\left(E_\nu,-P_\ell,\pi-\theta\right)
 =\frac{E_\ell\mp P_\ell}{2M{\mathcal R}}{\mathcal Z}, \\
  \rho_{+-}\left(E_\nu,P_\ell,\theta\right)
&=\rho_{-+}^*\left(E_\nu,P_\ell,\theta\right)
 =\frac{\ml\sin\theta}{4M{\mathcal R}}
  \left({\mathcal X}-i{\mathcal Y}\right)e^{i\varphi}.
\end{align*}
Here we have introduced the following notation:
\begin{align*}
{\mathcal X}=&\ \mp\left(2W_1-W_2-\frac{\ml^2}{M^2}W_4
               + \frac{E_\ell}{M}W_5\right)-\frac{E_\nu}{M}W_3, \\
{\mathcal Y}=&\ - \frac{P_\ell}{M}W_6, \\
{\mathcal Z}=&\   \left(1\pm\cos\theta\right)
                 \left(W_1\pm\frac{E_\nu\mp P_\ell}{2M} W_3\right) \\
             &\ +\frac{1\mp\cos\theta}{2}\left[W_2+\frac{E_\ell \pm P_\ell}{M}
  \left(\frac{E_\ell\pm P_\ell}{M}\,W_4-W_5\right)\right],
\end{align*}
and $\varphi=\varphi_+-\varphi_-$. Finally the projections of the
lepton polarization vector are given by
\begin{subequations}\label{PolarizationProjections}
\begin{equation}\label{P_PT}
\begin{pmatrix}
{\mathcal P}_P \\
{\mathcal P}_T
\end{pmatrix}
=\frac{\ml\sin\theta}{2M{\mathcal R}}
\begin{pmatrix}
 \cos\varphi & \sin\varphi \\
-\sin\varphi & \cos\varphi
\end{pmatrix}
\begin{pmatrix}
{\mathcal X} \\
{\mathcal Y}
\end{pmatrix},
\end{equation}
\begin{gather}
{\mathcal P}_L
 =\mp 1
  \pm\frac{\ml^2}{M^2{\mathcal R}}
     \left\{\left[\left(\dfrac{2M}{E_\ell+P_\ell}\right)W_1
  \pm\left(\frac{E_\nu-P_\ell}{E_\ell+P_\ell}\right)W_3\right]
     \cos^2\frac{\theta}{2}\right. \nonumber \\
   + \left.\left[\left(\frac{M}{E_\ell+P_\ell}\right)W_2
   + \left(\frac{E_\ell+P_\ell}{M}\right)W_4-W_5\right]\sin^2
     \frac{\theta}{2}\right\}.
\end{gather}
\end{subequations}
By putting $\varphi=0$\,,%
\footnote{We adopt this convention from here on.
          Therefore, according to Eq.~\eqref{P_PT},
          ${\mathcal P}_P\propto{\mathcal X}$ and
          ${\mathcal P}_T\propto{\mathcal Y}$.}
the formulas for ${\mathcal P}_P$ and ${\mathcal P}_L$ exactly
coincide with those of Ref.~\cite{Hagiwara:03} (obtained within
a noncovariant approach under assumption $W_6=0$).

Several simple conclusions immediately follow from
Eqs.~\eqref{PolarizationProjections}.
First, the perpendicular and transverse projections are
unobservable quantities in contrast with the longitudinal
projection of  $\boldsymbol{\mathcal{P}}$ and the degree of
polarization $\left|\boldsymbol{\mathcal{P}}\right|$.
Second, supposing that $W_6=0$ (as is the case in the Standard Model)
one can force the polarization vector to lie in the production
plane.
Third, a massless lepton is fully polarized,
$\boldsymbol{\mathcal{P}}=(0,0,\mp1)$.
In particular, at the energies of our interest, electron
is always fully polarized while in general,
this is not the case for muon and $\tau$ lepton.

We end this section with a short description of the
structure functions relevant to the three fundamental
subprocesses, DIS, QE and RES.

\subsection{Deep inelastic scattering (DIS)}
\label{sec:DIS}

In this case the relation between the structure functions
$W_n^{\text{(DIS)}}\left(x,Q^2\right)$ and measurable quantities
$F_n\left(x,Q^2\right)$ is obtained in a straightforward manner:
\begin{equation*}
W_1^{\text{(DIS)}}\left(x,Q^2\right) =
F_1\left(x,Q^2\right),\quad
W_n^{\text{(DIS)}}\left(x,Q^2\right) =
w^{-1}F_n\left(x,Q^2\right),\quad n=2,\dots,6.
\end{equation*}
Here $Q^2=-q^2$, $x=Q^2/2(pq)$ is the Bjorken scaling variable
and $w=(pq)/M^2$.

The generally accepted relations between the functions $F_1$,
$F_2$, $F_4$ and $F_5$ are
\begin{equation*}
F_1=F_5=\frac{F_2}{2x(1+R)}\left(1+\frac{x^2}{x'}\right),
\quad
F_4=\frac{1}{2x}\left(\frac{F_2}{2x}-F_1\right),
\end{equation*}
where $R$ is the ratio of longitudinal to transverse
cross sections in DIS and $x'=Q^2/(4M^2)$.

\subsection{Quasielastic scattering (QE)}
\label{sec:QES}

The charged hadronic currents describing the QE processes
are written \cite{LlewellynSmith:72}
\begin{align*}
\langle p,\,p'|\widehat{J}_\alpha^+|n,\,p\rangle&=
\cos\theta_C\overline{u}_p\left(p'\right)\varGamma_\alpha\,u_n(p), \\
\langle n,\,p'|\widehat{J}_\alpha^-|p,\,p\rangle&=
\cos\theta_C\overline{u}_n\left(p'\right)\overline{\varGamma}_\alpha\,u_p(p),
\end{align*}
where $\theta_C$ is the Cabibbo mixing angle, $p'=p+q$ and the vertex
function
\[
\varGamma_\alpha=\gamma_\alpha F_V
              +i\sigma_{\alpha\beta}\frac{q_\beta}{2M}F_M
              + \frac{q_\alpha}{M} F_S
              + \left(\gamma_\alpha F_A
              + \frac{p_\alpha+p'_\alpha}{M} F_T
              + \frac{q_\alpha}{M}F_P\right)\gamma_5,
\]
is defined through the six, in general complex, form factors
$F_i\left(q^2\right)$. A standard calculation then yields
\[
W_n^{\text{(QE)}}\left(x,Q^2\right) =
\cos^2\theta_Cw^{-1}\omega_n\left(Q^2\right)\delta(1-x),
\quad
n=1,\dots,6,
\]
where the functions $\omega_n$ are the bilinear combinations of the
form factors:
\begin{align*}
\omega_1 = &\   \left|F_A\right|^2+x'\left(\left|F_A\right|^2
              + \left|F_V+F_M\right|^2\right),                     \\
\omega_2 = &\   \left|F_V\right|^2+\left|F_A\right|^2
              + x'\left(\left|F_M\right|^2
              +4\left|F_T\right|^2\right), \\
\omega_3 = &\ -2\,\text{Re}\left[F_A^*\left(F_V+F_M\right)\right], \\
\omega_4 = &\  \text{Re}\left[F_V^*\left(F_S-\tfrac{1}{2}F_M\right)
                             -F_A^*\left(F_T+F_P\right)\right]
            + x'\left(\tfrac{1}{2}\left|F_M-F_S\right|^2
                                 +\left|F_T+F_P\right|^2\right)    \\
           &\ -\tfrac{1}{4}\left(1+x'\right)\left|F_M\right|^2
              +\left(1+\tfrac{1}{2}x'\right)\left|F_S\right|^2,    \\
\omega_5 = &\  2\,\text{Re}\left[F_S^*\left(F_V- x'F_M\right)
                                -F_T^*\left(F_A-2x'F_P\right)\right]
                                                        +\omega_2, \\
\omega_6 = &\  2\,\text{Im}\left[F_S^*\left(F_V- x'F_M\right)
                                +F_T^*\left(F_A-2x'F_P\right)\right].
\end{align*}
The only difference between this result and that
from Ref.~\cite{LlewellynSmith:72} is in the relative sign of the terms
in $\omega_6$.%
\footnote{According to Llewellyn Smith, the functions
          $\omega_5'=\omega_5-\omega_2$ and $\omega_6$ are,
          respectively, the real and imaginary parts of a unique
          function. Our examination does not confirm this property
          for the general case of nonvanishing second-class current
          induced form factors $F_S$ and $F_T$.
         }
Assuming all the form factors to be real we have $\omega_6=0$
and thus ${\mathcal P}_T=0$.

\subsection{Single resonance production (RES)}
\label{sec:RES}

Let us now consider the case of single $\Delta$ resonance
neutrinoproduction,
\[
         {\nu}_\ell+n(p)\to\ell^-+\Delta^+\left(\Delta^{++}\right),
\quad
\overline{\nu}_\ell+n(p)\to\ell^++\Delta^-\left(\Delta^{0 }\right).
\]
Assuming the isospin symmetry and applying the Wigner-Eckart theorem,
the hadronic weak current matrix elements are given by \cite{Hagiwara:03}
\begin{align*}
\langle\Delta^{+ },\,p'|\widehat{J}_\alpha|n,\,p\rangle &=
\langle\Delta^{0 },\,p'|\widehat{J}_\alpha|p,\,p\rangle  =
\cos\theta_C\overline{\psi}\mathstrut^{\,\beta}\left(p'\right)
\varGamma_{\alpha\beta}\,u(p), \\
\langle\Delta^{++},\,p'|\widehat{J}_\alpha|p,\,p\rangle &=
\langle\Delta^{- },\,p'|\widehat{J}_\alpha|n,\,p\rangle  =
\sqrt{3}\,
\cos\theta_C\overline{\psi}\mathstrut^{\,\beta}\left(p'\right)
\varGamma_{\alpha\beta}\,u(p).
\end{align*}
Here $\psi^\alpha\left(p'\right)$ is the Rarita-Schwinger spin-vector
for $\Delta$ resonance, $u(p)$ is the Dirac spinor for neutron or proton
and the vertex tensor $\varGamma_{\alpha\beta}$ is expressed in terms of
the eight weak transition form factors $C^{V,A}_{3,4,5,6}\left(Q^2\right)$
\cite{Schreiner:73}:
\begin{align*}
\varGamma_{\alpha\beta}
=&  \left[C^V_3\frac{g_{\alpha\beta} \widehat{q}-\gamma_\alpha q_\beta}{M  }
        + C^V_4\frac{g_{\alpha\beta}\left(qp'\right)-p'_\alpha q_\beta}{M^2}
        + C^V_5\frac{g_{\alpha\beta}\left(qp \right)-p _\alpha q_\beta}{M^2}
        + C^V_6\frac{q _\alpha q_\beta                                }{M^2}
    \right]\gamma_5 \\
 &  \!\!+ C^A_3\frac{g_{\alpha\beta} \widehat{q}-\gamma_\alpha q_\beta}{M  }
        + C^A_4\frac{g_{\alpha\beta}\left(qp'\right)-p'_\alpha q_\beta}{M^2}
        + C^A_5      g_{\alpha\beta}
        + C^A_6\frac{q_\alpha q_\beta}{M^2}.
\end{align*}
Below we will assume that time-reversal invariance holds so that all the
form factors are relatively real.
After accounting for the explicit form of a spin 3/2 projection operator
and computing the proper convolutions we arrive at the following expressions
for $W_n^{\text{(RES)}}$:
\begin{equation}\label{W_n_RES}
\begin{aligned}
W_n^{\text{(RES)}}&=
           \kappa\cos^2\theta_CMM_\Delta
           \left|\eta^{\Delta}_{\text{BW}}(W)\right|^2\!
           \sum\limits_{j,k=3\;\text{to}\;6}
           \left(V_n^{jk}C_j^V C_k^V+A_n^{jk}C_j^A C_k^A\right),
\quad n\ne3, \\
W_3^{\text{(RES)}}&=
           2\kappa\cos^2\theta_CMM_\Delta
           \left|\eta^{\Delta}_{\text{BW}}(W)\right|^2\!
           \sum\limits_{j,k=3\;\text{to}\;6}K_n^{jk}C_j^V C_k^A
\quad\text{and}\quad W_6^{\text{(RES)}}=0.
\end{aligned}
\end{equation}
Here $\kappa=2/3$ for $\Delta^{+ }$ and $\Delta^0$
production or $\kappa=2$ for $\Delta^{++}$ and $\Delta^-$
production;
\begin{equation*}\label{BW_D}
\left|\eta^{\Delta}_{\text{BW}}(W)\right|^2
=\frac{1}{2\pi}\left[\frac{\varGamma_\Delta(W)}
{\left(W^2-M_\Delta^2\right)^2+\varGamma_\Delta^2(W)/4}\right]
\end{equation*}
is the Breit-Wigner factor and $\varGamma_\Delta(W)$ is the running
width of the $\Delta$ resonance which can be estimated by assuming the
dominance of $S$-wave ($L=0$) or $P$-wave ($L=1$) $\Delta \to N\pi$
decay~\cite{Schreiner:73,Alvarez-Ruso:98}: 
\begin{equation*}\label{BW_Gamma}
\varGamma_\Delta(W) = \varGamma_\Delta^0
\left(\frac{M_\Delta}{W}\right)^L\left[\frac{p_\pi^\star(W)}
{p_\pi^\star(M_\Delta)}\right]^{2L+1},
\end{equation*}
where
\begin{equation*}\label{q_pi}
p_\pi^\star(W)=\sqrt{\left(\frac{W^2-M^2+m_\pi^2}{2W}\right)^2-m_\pi^2},
\end{equation*}
is the pion momentum in the rest frame of the $\Delta$ with invariant
mass $W=\left|p'\right|=\left|p+q\right|$; $M_\Delta$ and
$\varGamma_\Delta^0$ are, correspondingly, the central mass
and decay width of $\Delta$.

The coefficients $V_n^{jk}$, $A_n^{jk}$ and $K_n^{jk}$ are
found to be cubic polynomials in invariant dimensionless
variables $x$, $w$ and parameter $\zeta=M/M_\Delta$.
Only 70 among the total 144 coefficients are nonzero (see appendix).

\section{Single pion production in Rein--Sehgal model}
\label{sec:RS}

The Rein-Sehgal (RS) model~\cite{Rein:81} is one of the most
circumstantial and approved phenomenological tools for description
of single-pion production through baryon resonances in neutrino and
antineutrino interactions with nucleons.
It is incorporated into essentially all MC neutrino event generators
developed for both accelerator and astroparticle experiments and
is in agreement with available experimental data (see, e.g.,
Ref.~\cite{Zeller:03} for a recent review and further references).
However the RS model is not directly applicable to the $\nu_{\tau}$
and $\overline{\nu}_{\tau}$ induced reactions since it neglects
the final lepton mass. Due to the same reason, the model is not
suited for studying the lepton polarization.
In this section, we describe a simple generalization of the RS model
based upon the covariant form of the charged leptonic current $j_\lambda$
with definite lepton helicity discussed in Sect.~\ref{sec:PDM}
which allows us to take into account the final lepton mass and spin.

The charged hadronic current in the RS approach has been derived
in terms of the Feynman-Kislinger-Ravndal (FKR) relativistic quark
model \cite{Feynman:71} and its explicit form has been written in
the resonance rest frame (RRF); below we will mark this frame with
asterisk~(${}^\star$).
In RRF, the energy of the incoming neutrino, outgoing lepton,
target nucleon and the 3-momentum transfer are, respectively,
\begin{equation}\label{KinematicVariables} 
\begin{aligned}
E_\nu^\star  &= \frac{1}{2W}\left(2ME_\nu-Q^2-\ml^2\right)
              = \frac{E_\nu}{W}
                \left[M-\left(E_\ell-P_\ell\cos\theta\right)\right],    \\
E_\ell^\star &= \frac{1}{2W}\left(2ME_\ell+Q^2-\ml^2\right)
              = \frac{1}{ W}\left[ME_\ell
               +E_\nu\left(E_\ell-P_\ell\cos\theta\right)-\ml^2\right], \\
E_N^\star    &= W-\left(E_\nu^\star-E_\ell^\star\right)
              = \frac{M}{W}\left(M+E_\nu-E_\ell \right),                \\
\left|{\mathbf q}^\star\right| & =\frac{M}{W}\left|\mathbf{q}\right|
              = \frac{M}{W}
                \sqrt{E_\nu^2-2E_\nu P_\ell\cos\theta+P_\ell^2}.
\end{aligned}
\end{equation}
It is convenient to direct the spatial axes of RRF in such a way that
${\mathbf p}^\star=\left(0,0,-\left|\mathbf{q}^\star\right|\right)$
and $k_y^\star=k'_y{}^\star=0$.
These conditions lead to the following system of equations:
\begin{equation}\label{ki}
\left\{
\begin{aligned}
k_x^\star=k'_x{}^\star&=
\sqrt{\left(E_\nu^\star\right)^2-\left(k_z^\star\right)^2}, \\
k_z^\star-k'_z{}^\star&=\left|\mathbf{q}^\star\right|, \\
k_z^\star+k'_z{}^\star&=
\frac{1}{\left|\mathbf{q}^\star\right|}\left[\left(E_\nu^\star\right)^2
-\left(E_\ell^\star\right)^2+\ml^2\right].
\end{aligned}
\right.
\end{equation}
By using Eqs.~\eqref{KinematicVariables} and \eqref{ki}
we find the components of the lepton spin 4-vector:
\begin{align*}
s_0^\star&=\frac{1}{\ml W}
\left[M P_\ell+E_\nu\left(P_\ell-E_\ell\cos\theta\right)\right], \\
s_x^\star&=\frac{E_\nu E_\ell}{\ml\left|\mathbf{q}\right|}\sin\theta, \\
s_y^\star&=0, \\
s_z^\star&=\frac{1}{\ml\left|\mathbf{q}\right|W}
           \left[\left(E_\nu\cos\theta-P_\ell\right)
           \left(ME_\ell-\ml^2+E_\nu E_\ell\right)
           -E_\nu P_\ell\left(E_\nu-P_\ell\cos\theta\right)\right].
\end{align*}
Then, by applying the general equations~\eqref{j_components}, the
components of the leptonic current in RRF with the lepton helicity
$\lambda$ measured in lab.\ frame, are expressed as
\begin{align*}
j_0^\star &= N_\lambda\ml\frac{E_\nu}{W}(1-\lambda\cos\theta)
               \left(M-E_\ell-\lambda P_\ell\right), \\
j_x^\star &= N_\lambda\ml\frac{E_\nu}{\left|\mathbf{q}\right|}\sin\theta
               \left(P_\ell-\lambda E_\nu\right),    \\
j_y^\star &=  i\lambda N_\lambda\ml E_\nu\sin\theta, \\
j_z^\star &= N_\lambda\ml\frac{E_\nu}{\left|\mathbf{q}\right|W}(1-\lambda\cos\theta)
               \left[\left(E_\nu+\lambda P_\ell\right)\left(M-E_\ell\right)
            +P_\ell\left(\lambda E_\nu+2E_\nu\cos\theta-P_\ell\right)\right].
\end{align*}
On the other hand, in the spirit of the RS model, the leptonic current
still can be treated as the intermediate $W$ boson polarization 4-vector.
Therefore it may be decomposed into three polarization 4-vectors
corresponding to left-handed, right-handed and scalar polarization:
\begin{gather*}
j^\alpha_\lambda=K^{-1}\left[
c_L^\lambda e^\alpha_{L}+
c_R^\lambda e^\alpha_{R}+
c_S^\lambda e^\alpha_{(\lambda)}
\right], \\
\begin{aligned}
e^\alpha_L &= \frac{1}{\sqrt{2}}(0, 1,-i,0), \\
e^\alpha_R &= \frac{1}{\sqrt{2}}(0,-1,-i,0), \\
e^\alpha_{(\lambda)} &=
\frac{1}{\sqrt{Q^2}}
\left(\Q^\star_{(\lambda)},0,0,\nu^\star_{(\lambda)}\right).
\end{aligned}
\end{gather*}
Here the vectors $e^\alpha_L$ and $e^\alpha_R$ are the same as in
Ref.~\cite{Rein:81} while $e^\alpha_{(\lambda)}$ has been modified
to include the lepton mass effect:
\begin{equation*}
\Q^\star_{(\lambda)}=\frac{K\sqrt{Q^2}}{c_S^\lambda}j_0^\star,
\quad
\nu^\star_{(\lambda)}=\frac{K\sqrt{Q^2}}{c_S^\lambda}j_z^\star,
\quad
K=\frac{\left|\mathbf{q}\right|}{E_\nu\sqrt{2Q^2}}.
\end{equation*}
Then the coefficients $c_i^{\lambda}$ are explicitly defined through
the components $j_\lambda^\alpha$ in RRF as
\begin{equation*}
c^\lambda_L=\frac{K}{\sqrt{2}}
\left(j_x^\star+ij_y^\star\right),
\quad
c^\lambda_R=-\frac{K}{\sqrt{2}}
\left(j_x^\star-ij_y^\star \right),
\quad
c^\lambda_S=K\sqrt{|(j_0^\star)^2-(j_z^\star)^2|}.
\end{equation*}
For antineutrino induced reactions one have to take into account
relation \eqref{Nu-Nubar_relation}. It yields
\begin{gather*}
  \left[\mathcal{Q}^{\star}_{( \lambda)}\right]_{\overline{\nu}}
= \left[\mathcal{Q}^{\star}_{(-\lambda)}\right]_{\nu},
\quad
  \left[\nu^{\star}_{( \lambda)}\right]_{\overline{\nu}}
= \left[\nu^{\star}_{(-\lambda)}\right]_{\nu}, \\
         \left[c^{ \lambda}_L\right]_{\overline{\nu}}
= \lambda\left[c^{-\lambda}_R\right]_{\nu},
\quad
         \left[c^{ \lambda}_R\right]_{\overline{\nu}}
= \lambda\left[c^{-\lambda}_L\right]_{\nu},
\quad
          \left[c^{ \lambda}_S\right]_{\overline{\nu}}
= -\lambda\left[c^{-\lambda}_S\right]_{\nu}.
\end{gather*}

Within the extended RS model, the elements of the polarization
density matrix may be written as the superpositions of the
partial cross sections%
\footnote{For the reader's convenience we use the same definitions
          and (almost) similar notation as in Ref.~\cite{Rein:81}.}
$\sigma_L^{\lambda\lambda'}$,
$\sigma_R^{\lambda\lambda'}$ and
$\sigma_S^{\lambda\lambda'}$:
\begin{equation*}
\rho_{\lambda\lambda'}=\frac{\varSigma_{\lambda\lambda'}}
{\varSigma_{++}+\varSigma_{--}},
\quad
\varSigma_{\lambda\lambda'}=\sum\limits_{i=L,R,S}\!
c_i^{\lambda}c_i^{\lambda'}\sigma_i^{\lambda\lambda'},
\end{equation*}
and the differential cross section is given by
\begin{equation*}
\frac{d^2\sigma}{dQ^2dW^2}=
\frac{G_F^2\cos^2\theta_CQ^2}{2\pi^2M\left|\mathbf{q}\right|^2}
\left(\varSigma_{++}+\varSigma_{--}\right).
\end{equation*}
The partial cross sections are found to be the bilinear superpositions of
the reduced amplitudes for producing a $N\pi$ final state with allowed
isospin by a charged isovector current:
\begin{align*}
\sigma_{L,R}^{\lambda\lambda'}&=\frac{\pi W}{2M}\left(
A_{\pm 3}^{\lambda}A_{\pm3}^{\lambda'}+
A_{\pm 1}^{\lambda}A_{\pm1}^{\lambda'}\right), \\
\sigma_{S}^{\lambda\lambda'}&=\frac{\pi M\left|\mathbf{q}\right|^2}{2WQ^2}
\left(A_{0+}^{\lambda}A_{0+}^{\lambda'}+
      A_{0-}^{\lambda}A_{0-}^{\lambda'}\right).
\end{align*}
The amplitudes for neutrino induced reactions are
\begin{align*}
A^\lambda_\varkappa\left(p\pi^+\right) &=
 \sqrt{3}           \sum_{(I=3/2)}a^\lambda_\varkappa\left(\N_3^{\ast}\right), \\
A^\lambda_\varkappa\left(p\pi^0\right) &=
 \sqrt{\tfrac{2}{3}}\sum_{(I=3/2)}a^\lambda_\varkappa\left(\N_3^{\ast}\right)
-\sqrt{\tfrac{1}{3}}\sum_{(I=1/2)}a^\lambda_\varkappa\left(\N_1^{\ast}\right), \\
A^\lambda_\varkappa\left(n\pi^+\right) &=
 \sqrt{\tfrac{1}{3}}\sum_{(I=3/2)}a^\lambda_\varkappa\left(\N_3^{\ast}\right)
+\sqrt{\tfrac{2}{3}}\sum_{(I=1/2)}a^\lambda_\varkappa\left(\N_1^{\ast}\right).
\end{align*}
Here $\varkappa=\pm3,\;\pm1,\;0\pm$ and only those resonances are allowed
to interfere which have the same spin and orbital angular momentum as is
in the following typical example describing the $n\pi^+$ final state:
\begin{align*}
3A_\varkappa^{\lambda}&\left(n\pi^+\right)
 A_\varkappa^{\lambda'}\left(n\pi^+\right)=                           \\
&\ \left[\sum a^{\lambda }_\varkappa\left(S^+_{31}\right)
     +\sqrt{2}\sum a^{\lambda }_\varkappa\left(S^+_{11}\right)\right]\!
\left[\sum a^{\lambda'}_\varkappa\left(S^+_{31}\right)
     +\sqrt{2}\sum a^{\lambda'}_\varkappa\left(S^+_{11}\right)\right] \\
&\   +\!\sum_{j=1,3}
\left[\sum a^{\lambda}_\varkappa\left(P^+_{3j}\right)
     +\sqrt{2}\sum a^{\lambda }_\varkappa\left(P^+_{1j}\right)\right]\!
\left[\sum a^{\lambda'}_\varkappa\left(P^+_{3j}\right)
     +\sqrt{2}\sum a^{\lambda'}_\varkappa\left(P^+_{1j}\right)\right] \\
&\   +\!\sum_{j=3,5}
\left[\sum a^{\lambda}_\varkappa\left(D^+_{3j}\right)
     +\sqrt{2}\sum a^{\lambda }_\varkappa\left(D^+_{1j}\right)\right]\!
\left[\sum a^{\lambda'}_\varkappa\left(P^+_{3j}\right)
     +\sqrt{2}\sum a^{\lambda'}_\varkappa\left(P^+_{1j}\right)\right] \\
&\   +\!\sum_{j=5,7}
\left[\sum a^{\lambda}_\varkappa\left(F^+_{3j}\right)
     +\sqrt{2}\sum a^{\lambda }_\varkappa\left(F^+_{1j}\right)\right]\!
\left[\sum a^{\lambda'}_\varkappa\left(P^+_{3j}\right)
     +\sqrt{2}\sum a^{\lambda'}_\varkappa\left(P^+_{1j}\right)\right].
\end{align*}

Any amplitude $a^\lambda_\varkappa\left(\N_\imath^{\ast}\right)$ referring to
one single resonance $\N_\imath^{\ast}$ in a definite state of isospin, charge
and helicity consists of two factors which describe the production and
subsequent decay of the resonance:
\[
a^\lambda_\varkappa\left(\N^\ast_\imath\right)=
f^\lambda_\varkappa\left(\nu\N\to\N^\ast_\imath\right)\,
\eta(\N^{\ast}_\imath\to\N\pi) \equiv
f^{\lambda(\imath)}_\varkappa\,\eta^{(\imath)}.
\]
The decay amplitudes, $\eta^{(\imath)}$, can be split into three factors,
\[
\eta^{(\imath)}=\mathrm{sign}(\N^{\ast}_\imath)\,
\sqrt{\chi_\imath}\,\eta^{(\imath)}_{\text{BW}}(W),
\]
irrespective of isospin, charge or helicity of the resonance.
Here, the first factor is the decay sign for resonance $\N^\ast_\imath$
(see Table~III of Ref.~\cite{Rein:81}), $\chi\strut_\imath$ is the elasticity
of the resonance taking care of the branching ratio into the ${\pi}N$ final
state and $\eta^{(\imath)}_{\text{BW}}(W)$ is the properly normalized
Breit-Wigner term with the running width specified by the ${\pi}N$ partial
wave from which the resonance arises (Eq.~(2.31) in Ref.~\cite{Rein:81}).

The resonance production amplitudes, $f^{\lambda(\imath)}_\varkappa$,
can be calculated within the FKR quark model in exactly the same way as in
Ref.~\cite{Rein:81}. It can be shown that they have the same structure
as that given in Table~II of Ref.~\cite{Rein:81} with the only important
difference: the three coefficient functions $S$, $B$ and $C$ involved into
the definitions of the amplitudes have to be modified. Indeed, since in our
approach the structure of the polarization 4-vector $e^\alpha_{(\lambda)}$
has been changed with respect to that of the original RS model (by including
the lepton mass and spin), we have to recalculate its inner products with
the vector and axial hadronic currents. To do this, we used the explicit
form for the FKR currents given by Ravndal~\cite{Ravndal:73}. As a result,
the coefficients $S$, $B$ and $C$ (and thus the resonance production
amplitudes) become parametricaly dependent of the lepton mass and helicity:
\begin{align*}
S=S^V =&\   \left(\nu^\star_{(\lambda)}\nu^\star
           -\Q^\star_{(\lambda)}\left|\mathbf{q}^\star\right|\right)
            \left(1+\frac{Q^2}{M^2}-\frac{3W}{M}\right)
            \frac{G^V\left(Q^2\right)}{6\left|\mathbf{q}\right|^2},       \\
B=B^A =&\   \sqrt{\frac{\Omega}{2}}
            \left(\Q^\star_{(\lambda)}+\nu^\star_{(\lambda)}
            \frac{\left|\mathbf{q}^\star\right|}{aM}\right)
            \frac{ZG^A\left(Q^2\right)}{3W\left|\mathbf{q}^\star\right|}, \\
C=C^A =&\   \left[\left(\Q^\star_{(\lambda)}\left|\mathbf{q}^\star\right|
           -\nu^\star_{(\lambda)}\nu^\star\right)
            \left(\frac{1}{3}+\frac{\nu^\star}{aM}\right)\right.          \\
       &\   \left.+\nu^\star_{(\lambda)}\left(\frac{2}{3}W
           -\frac{Q^2}{aM}+\frac{n\Omega}{3aM}\right)\right]
            \frac{ZG^A\left(Q^2\right)}{2W\left|\mathbf{q}^\star\right|}.
\end{align*}
Here
\[
\nu^\star=E_\nu^\star-E_\ell^\star=\frac{M\nu-Q^2}{W},
\quad
a=1+\frac{W^2+Q^2+M^2}{2MW},
\]
$G^{V,A}\left(Q^2\right)$ are the vector and axial transition form factors and
the remaining notation is explained in Ref.~\cite{Rein:81}. Other 5 coefficients
listed in Eq.~(3.11) of Ref.~\cite{Rein:81} are left unchanged.

\section*{Acknowledgments}

The authors thank S.~M.~Bilenky, A.~V.~Efremov, I.~F.Ginzburg, D.~V.~Naumov and
O.~V.~Teryaev for useful conversations.
V.~L. and K.~K. are very grateful to the Physics Department of
Florence University for warm hospitality and financial
support during the final stage of this work.

\appendix
\section*{Appendix: Coefficients $V_i^{jk}$, $A_i^{jk}$, $K_i^{jk}$.}

Here we list the seventy \emph{nonzero} coefficients $V_i^{jk}$,
$A_i^{jk}$ and $K_i^{jk}$ appearing in Eqs.~\eqref{W_n_RES}.
\begin{align*}
V^{33}_1 &=  \zeta^3(1-2x)^2w^3+\zeta[1+\zeta^2(1-2x)^2]      w^2
                                                    +2(1+\zeta)xw, \\
V^{34}_1 &=  \zeta^2(1-2x)^2w^3+[1-\zeta(2-\zeta)(1-2x)](1-2x)w^2, \\
V^{35}_1 &=  \zeta^2(1-2x)  w^3+[1-\zeta(2-\zeta)(1-2x)]      w^2, \\
V^{44}_1 &=  \zeta  (1-2x)^2w^3        -(1-\zeta)(1-2x)^2     w^2, \\
V^{45}_1 &= 2\zeta  (1-2x)  w^3        -(1-\zeta)(1-4x)       w^2, \\
V^{55}_1 &=  \zeta          w^3        -(1-\zeta)             w^2;
\end{align*}
\begin{align*}
V^{33}_2 &= 2\zeta^3      xw^2+2\zeta(1+\zeta^2)xw, \\
V^{34}_2 &= 2\zeta^2      xw^2+2     (1-\zeta)^2xw, \\
V^{35}_2 &= 2\zeta^2(1+2x)xw^2+2     (1-\zeta)^2xw, \\
V^{44}_2 &= 2\zeta        xw^2-2     (1-\zeta)  xw, \\
V^{45}_2 &= 4\zeta        xw^2-4     (1-\zeta)  xw, \\
V^{55}_2 &= 4\zeta^3x^2w^3+2\zeta[1-2\zeta(1-\zeta)x]xw^2-2(1-\zeta)xw;
\end{align*}
\begin{align*}
V^{33}_4 &= 2\zeta^3(1- x)w^2+2\zeta^3             (1-x)w-1-\zeta, \\
V^{34}_4 &= 2\zeta^2(1- x)w^2+[1-2\zeta  (2-\zeta)(1-x)]w        , \\
V^{35}_4 &= 2\zeta^2      w^2-    \zeta  (2-\zeta)      w        , \\
V^{36}_4 &=-2\zeta^2     xw^2-           (1-\zeta)^2    w        , \\
V^{44}_4 &= 2\zeta  (1- x)w^2-   2       (1-\zeta)(1-x) w        , \\
V^{45}_4 &= 2\zeta        w^2-   2       (1-\zeta)      w        , \\
V^{46}_4 &=-2\zeta        w^2+   2       (1-\zeta)      w        , \\
V^{55}_4 &=  \zeta^3      w^3-    \zeta^2(1-\zeta)      w^2      , \\
V^{56}_4 &= 2\zeta^3(1-2x)w^3
           -2\zeta\left[1+\zeta(1-\zeta)(1-2x)   \right]w^2
                                            +2(1-\zeta) w        , \\
V^{66}_4 &=  \zeta^3(1-2x)^2w^3
           - \zeta\left[\zeta(1-\zeta)(1-2x)^2-2x\right]w^2
                                            -2(1-\zeta)xw;
\end{align*}
\begin{align*}
V^{33}_5 &= 2\zeta^3         w^2       +2\zeta(1+\zeta^2) w, \\
V^{34}_5 &= 2\zeta^2         w^2       +2(1-\zeta)^2      w, \\
V^{35}_5 &= 2\zeta^2(1+2x)   w^2       +2(1-\zeta)^2      w, \\
V^{36}_5 &=-4\zeta^2      x^2w^2       -2(1-\zeta)^2x     w, \\
V^{44}_5 &= 2\zeta           w^2       -2(1-\zeta)        w, \\
V^{45}_5 &= 4\zeta           w^2       -4(1-\zeta)        w, \\
V^{46}_5 &=-4\zeta        x  w^2       +4(1-\zeta)  x     w, \\ 
V^{55}_5 &= 4\zeta^3      x  w^3
           +2\zeta[1-2\zeta(1-\zeta)x]w^2-2(1-\zeta)w      , \\
V^{56}_5 &= 4\zeta^3(1-2x)x  w^3
           -4\zeta[\zeta(1-\zeta)(1-2x)+1]xw^2+4(1-\zeta)xw;
\end{align*}
\begin{align*}
A^{33}_1 &=  \zeta^3(1-2x)^2w^3+\zeta[1+\zeta^2(1-2x)^2]w^2
                                            -2(1-\zeta)xw  , \\
A^{34}_1 &=  \zeta^2(1-2x)^2w^3
                         +[1+\zeta(2+\zeta)(1-2x)](1-2x)w^2, \\
A^{35}_1 &=  \zeta^2(1-2x)  w^2+[1+\zeta(2+\zeta)(1-2x)]w  , \\
A^{44}_1 &=  \zeta  (1-2x)^2w^3+(1+\zeta)(1-2x)^2       w^2, \\
A^{45}_1 &= 2\zeta  (1-2x)  w^2+2(1+\zeta)(1-2x)        w  , \\
A^{55}_1 &=  \zeta          w  +1+\zeta;
\end{align*}
\begin{align*}
A^{33}_2 &= 2\zeta^3xw^2+2\zeta(1+\zeta^2)xw, \\
A^{34}_2 &= 2\zeta^2xw^2+2     (1+\zeta)^2xw, \\
A^{35}_2 &= 2\zeta^2xw,                       \\
A^{44}_2 &= 2\zeta  xw^2+2     (1+\zeta)  xw, \\
A^{55}_2 &=  \zeta^3 w+\zeta^2(1+\zeta);
\end{align*}
\begin{align*}
A^{33}_4 &= 2\zeta^3(1-x)   w^2+2\zeta^3(1-x)w+1-\zeta   , \\
A^{34}_4 &= 2\zeta^2(1-x)   w^2+[1+2\zeta(2+\zeta)(1-x)]w, \\
A^{35}_4 &= 2\zeta^2        w  +\zeta(2+\zeta)           , \\
A^{36}_4 &=-2\zeta^2x       w^2- (1+\zeta)^2   w         , \\
A^{44}_4 &= 2\zeta  (1-x)   w^2+2(1+\zeta)(1-x)w         , \\
A^{45}_4 &= 2\zeta          w  +2(1+\zeta)               , \\
A^{46}_4 &=-2\zeta          w^2-2(1+\zeta)w              , \\
A^{55}_4 &=  \zeta^3        w  +\zeta^2(1+\zeta)         , \\
A^{56}_4 &= 2\zeta^3(1-2x)  w^2
           +2\zeta[\zeta(1+\zeta)(1-2x)-1]w-2(1+\zeta)   , \\
A^{66}_4 &=  \zeta^3(1-2x)^2w^3
           + \zeta[\zeta(1+\zeta)(1-2x)^2+2x]w^2+2(1+\zeta)xw;
\end{align*}
\begin{align*}
A^{33}_5 &= 2\zeta^3        w^2+2\zeta  (1+\zeta^2)     w, \\
A^{34}_5 &= 2\zeta^2        w^2+2       (1+\zeta)^2     w, \\
A^{35}_5 &= 2\zeta^2(1+x)   w  +        (1+\zeta)^2      , \\
A^{36}_5 &=-4\zeta^2x^2     w^2-2       (1+\zeta)^2    xw, \\
A^{44}_5 &= 2\zeta          w^2+2       (1+\zeta)       w, \\
A^{45}_5 &= 2\zeta          w  +2       (1+\zeta)        , \\
A^{46}_5 &=-4\zeta  x       w^2-4       (1+\zeta)      xw, \\
A^{55}_5 &= 2\zeta^3        w  +2\zeta^2(1+\zeta)        , \\
A^{56}_5 &= 2\zeta^3(1-2x)  w^2+2\zeta^2(1+\zeta)(1-2x) w;
\end{align*}
\begin{align*}
K^{33}_3 &=-2\zeta^3(1-2x)^2w^2+2\zeta(2-3x)    w, \\
K^{34}_3 &=- \zeta^2(1-2x)^2w^2+2(1+\zeta)(1-2x)w, \\
K^{35}_3 &=- \zeta^2(1-2x)  w  +2(1+\zeta)       , \\
K^{43}_3 &=- \zeta^2(1-2x)^2w^2+2(1-\zeta)(1-2x)w, \\
K^{44}_3 &=  \zeta  (1-2x)^2w^2                  , \\
K^{45}_3 &=  \zeta  (1-2x)  w                    , \\
K^{53}_3 &=- \zeta^2(1-2x)  w^2-2(1-\zeta)      w, \\
K^{54}_3 &=  \zeta  (1-2x)  w^2                  , \\
K^{55}_3 &=  \zeta          w                    .
\end{align*}

\end{document}